# A comparative study of the behaviour of forsterite melts under atmospheric and sub-atmospheric conditions

*Biswajit Mishra[1], Prateek Manvar[1], Kaushik Choudhury[2], S. Karagadde[1] and Atul Srivastava[1,1]*

[1]*Department of Mechanical Engineering, Indian Institute of Technology Bombay, Powai, Mumbai 400076, India*
[2]*School of Applied Sciences, KIIT, Bhubaneshwar, Odisha 751024, India*

## Abstract

The study is focussed towards understanding the difference in behaviour of forsterite droplets subjected to sub-atmospheric conditions. Melt droplets, 1.5-2.5 mm in diameter, are made to crystallize under levitated conditions. The spherule is initially superheated by about 250ºC above its liquids temperature, for both atmospheric and sub-atmospheric conditions, and then, subsequently cooled at different cooling rates. Crystallization of molten droplets was observed from their hypercooled states in all the cases. It was found that the level of undercooling was more (~150K-200K) for sub-atmospheric cases. In addition, the degree of recalescence was also found to be higher. However, the varying cooling rates did not produce any considerable effect on the level of undercooling or, even, the degree of recalescence. For better insight into the dynamics of crystallization, in situ visualization of the recalescence process was made possible by use of high-speed camera. A clear difference in the growth mechanisms was observed. The role of cooling rate in affecting crystallization was seen as the difference in the growth of the crystal-front; from a clear rim front, in high cooling rates, to a faded irregularly shaped crystal front, in low cooling rates. Two proper crystal fronts were observed in all the cases, which correspond to surface and volumetric crystallizations. It was observed that the molten droplets, under sub-atmospheric conditions, undergo high rate of volumetric crystallization, which is marked by the sightings of surface textures during the recalescence process. High-speed camera images of recalescence provided direct observational proof of the delay in volumetric crystallization, compared to surface crystallization.

---

[1] Corresponding author: Dr. Atul Srivastava
Professor, Department of Mechanical Engineering
IIT Bombay, Powai, Mumbai- 400076, India
Tel: +91-22-25767531; Fax: +91-22-25726875; +91-22-25723480
Email: atulsr@iitb.ac.in; atuldotcom@gmail.com

## 1. Introduction

Forsterite, largely found in chrondites, has always been considered as one of the most primitive metal siicates. It has been found in the cometary dust of the meterorites [1] and has been observed as tiny crystals in the dusty clouds of gas around a forming star [2]. The researchers seek to get an insight into the dynamics of evolution of materials, from the protoplanetary disk, into the present-day form [3], by understanding the nature of crystallization of the chondrites. Apart from this, a conceptual understanding of the behaviour of the materials formed from high temperature crystallization helps the semiconductor industry in the development of new material designs [4][5].

However, many models have been developed explaining the crystallization of chondrules in the early solar system [6][7][8]. One section of the astronomical society argues that crystallization took place under low cooling rates, and because of it a considerable number of experiments have been performed under such conditions [9][10][11] with an intent to reproduce the same textures as observed [12], while the other section believes crystallization took place under rapid cooling rates [13]. Nagashima et al [14], however, justified that under low cooling rates (~1000K/hr), the collison time interval between the cosmic dust and the chondrule melt would be around 30-39mins, which is too long considering the high density of cosmic dust available around the chondrule formation region. This, otherwise, suggested that in order to have very short collison time interval, the cooling rate should be in orders of 100K/s. Tsukamoto et al. [15] tried for the first time in reproducing the chondrule textures using containerless experimental conditions (aero-acoustic levitator). Tangeman et al. [16] used aero-acoustic levitation in synthesis of vitreous forsterite. It was observed that slow cooling of the levitated forsterite melt lead to recalescence, ejecting out the latent heat of crystallization which leads to increase in the melt's temperature [12], while the rapid cooling lead to formation of glassy forsterite.

With the recent developments, researchers have made efforts into understanding the crystallization process by use of high-end scientific tools in situ. Aoyama and Kuribayashi [4] used high speed camera to visualize and investigate the growth mechanism processs of Si and Ge. Srivastava et al. [17] developed in-house optical setup (a combination of shadowgraph and schlieren techniques) for in situ visualization of convections inside the levitated forsterite melt before and during the crystallization. Nagashio et al. [18] performed real-time XRD experiments to understand the grain refinement dynamics involved during the solidification of Si. These growth mechanisms have been observed to vary with different levels of undercoolings for Si and Ge melts [19]. A proper motivated numerical study was done by Miura et al. [20] on the effect of supercooling and cooling rate on the growth mechanism for forsterite melts. In order to varying different levels of undercooling, the melts have to be heterogeneously nucleated. However, the study was based on heterogenous nucleation and its growth kinetics are completely different from the homogenous case.

Although the containerless method best suites the objective of experimentally simulating the chondule formation process, but the case of homogenous or heterogenous crystallization of chondrules during their formation still remains an open debate. In addition to this, a proper clarification on the nature of the surrounding medium during chondrule formation has not been reported. Many models suggests the formation of chondrules under different magnitudes of gravity [21][13]. Although the most likely scenario is the case of proper vacuum conditions with zero gravity, but the production of both the conditions simultaneously in lab environment is next to impossible. Irrespectively, Liu et al [22] investigated the effect of microgravity conditions on the nucleation rate of $CaCO_3$ from an aqueous solution. The nucleation rate was found to increase, by four order of magnitude, under microgravity conditions. Nagashima et al. [14] found that nucleation of enstatite melts was delayed under microgravity conditions. However, no study has been reported on the effect of vacuum conditions on the melts' behaviour. Moreover, the studies performed under microgravity conditions essentially involved heterogenous nucleation cases; which shed no light of the microgravity effect on the other aspect of chondrule formation, i.e. homogenous nucleation case.

Against this backdrop, the current study is, thus, aimed at to gain insight into the effect of pressure conditions (atmospheric and sub-atmospheric) on the growth mechanisms of a homogenously nucleated forsterite melt. In order to ensure homogenous nucleation, aerodynamic levitation system was used in the experiments. The sub-atmospheric pressure condition was set at 50mBar using a turbo-vacuum pump and a comparative study was conducted between atmospheric and sub-atmospheric condition. In addition, the effect of cooling rate (100K/s, 200K/s and 400K/s) on the melt's behaviour has also been investigated. For proper understanding of the growth kinetics, high speed camera was used. A discussion from the thermal perspective has been made on the difference observed. However, lastly, SEM images have been provided in support of the claims made in comparisons.

## 2. Instrumentation and Experimental Procedure

The present study was carried on an aerodynamic High Temperature Conical Nozzle Levitator (HTCNL), developed by MDI (USA). A chondrule of Mg salt was levitated on a gas jet and heated to the temperatures at which it was completely melt and then was cooled at different cooling rates to understand the effect of cooling on the surface morphology. The studies were conducted under two different conditions – at atmospheric pressure and at sub-atmospheric pressure. The system was integrated with various diagnostics to study the cooling process and to understand the effect it has on the surface morphology. The complete set-up (including the levitator and the diagnostics) has been depicted in Figure 1.

The major components of the HTCNL system are, levitation chamber fitted with glass windows, $CO_2$ laser, vacuum pump, digital pyrometer and high speed camera. The $CO_2$ laser (Firestar i400) along with compatible controller is capable of delivering a power of 400W on the chondrule surface with the help of beam steering and focussing optics fitted with it. The $CO_2$ laser emits a t a wavelength of 10.6 micrometre and falls under class IV owing to its power output and damage potential. However, in our case the laser beam being guided through mirrors and lenses is so enclosed, using tubing of anodized aluminium that the laser is safe for operation and is considered class I for all practical applications. Also there are appropriate interlocks that switched off the laser radiation if any of the viewports are open in the levitation chamber. The levitation chamber, as shown in the figure, is made up of stainless steel and has viewports (NW 25 and NW 40) for imaging the levitated chondrule and connecting vacuum lines. This chamber houses the conical nozzle and the base of the nozzle is connected to the gas inlet, through which the gas can be pumped to achieve aerodynamic levitation. The entire levitation chamber, including, the gas inlet is maintained at $20^0C$ by circulating chilled water. The laser temperature is maintained static at $20^0C$ (so as to have a steady power output) using another chilling unit that takes away the heat and maintains the laser temperature. The third most important part of the system was high speed camera (make………) capable of capturing images at the rate of few thousandds of frames per second. This camera enables the observation of minute changes taking place on the surface of the chondrule while cooling and subsequent crystallisation takes place. However, at such high speeds the exposure times are very low and lead to poor image quality. To circumvent this, an additional light source was used to illuminate the chondrule. To conduct the experiments under sub-atmospheric conditions, a vacuum pump (make Edwards, dry scroll) was employed that has a peak pumping capacity of ………………………. In addition a Pirani gauge (make ………………….) was used to measure the pressure in the chamber. The vacuum gage was calibrated using standard procedures before conducting each set of the experiments. The real-ime temperature measurement of the chondrule was performed using a digital pyrometer (make…………..), which is capable of sensing temperature based-on the radiation coming out of the heated object, with a fixed emissivity, within a range of 540 – 3000K. The schematic description of the levitation and imaging has been given in Figure 2.

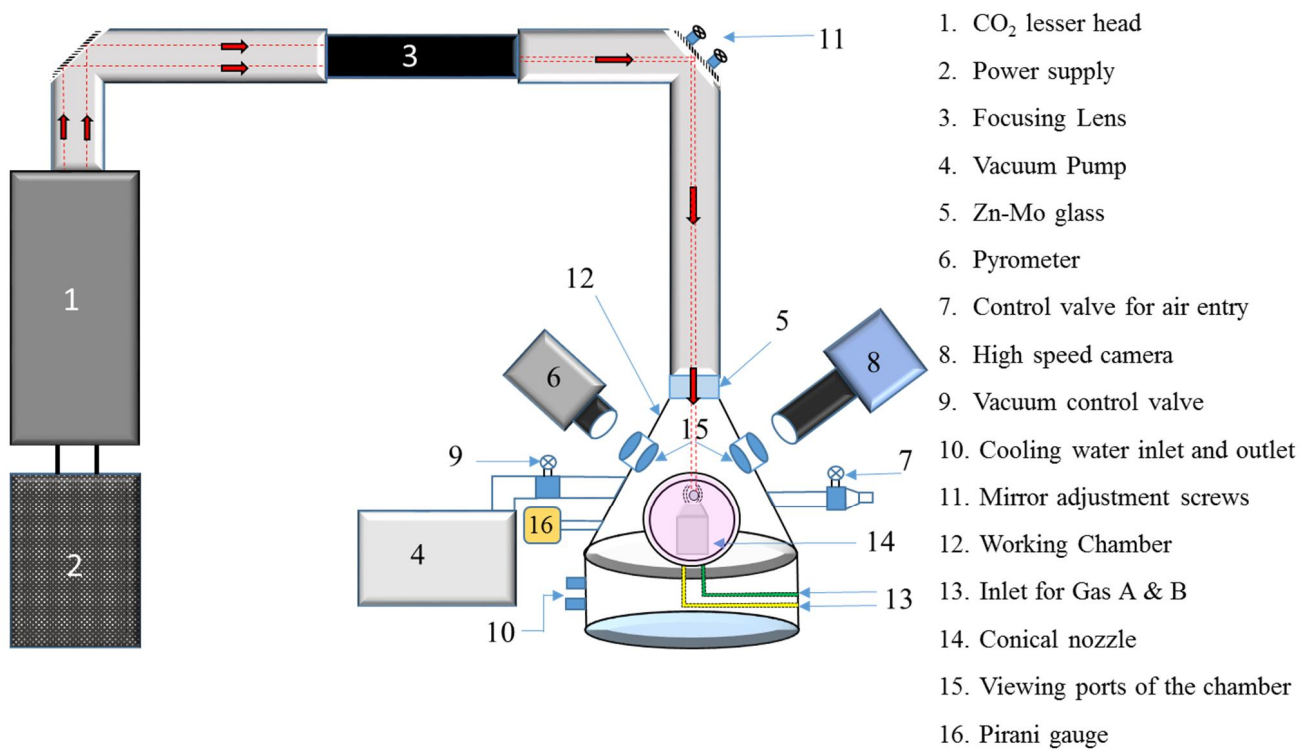

Figure 1: Schematic of the whole experimental setup.

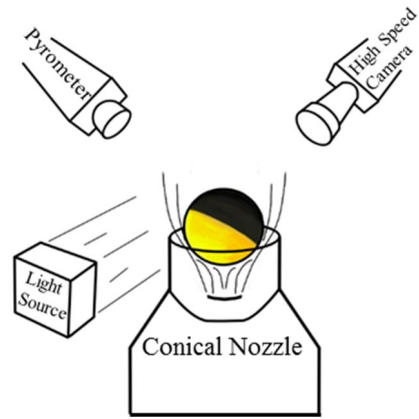

Figure 2: Schematic of the important accessories involved during an experiment

In order to maintain uniformity in the results, each chondrule was prepared by carefully weighing forsterite powder sample on a Mettler Toledo digital balance. The weighed powder was placed on a copper hearth, and was exposed to the laser radiation. The radiation meleted the powder and subsequently spherical chondrules were formed under the high surface tension of the melt. The sizes of the chondrules formed were measured and was found to be in the range 2±0.3mm. The chondrules so formed were held using vacuum-tweezers and were placed, one at a time, on the nozzle of the HTCNL for performing the experiments. The stability of suspended chondrule melt and the constancy of the initial temperature for all the experiments was maintained by critically adjusting both the laser power and the gas flow rate. For all the experiments reported in this article the levitation was achieved using argon gas. The stable levitation of the chondrule became a bigger challenge when the experiments were performed in the sub-atomic conditions. The chondrule, owing to its small mass, tended to move along the air stream created by the vacuum pump and hence would flew off the nozzle if proper care is not exercised. To achieve this, a leak was introduced in the chamber by fitting one of the ports (diametrically opposite to the suction port) with a nozzle valve. Initially, the vacuum pump was run at full capacity while the nozzle valve remained fully open. This was like levitation under atmospheric condition. Once this was achieved, the nozzle valve was closed slowly while controlling the incoming gas flow and the laser power. The gas inlet was controlled so as to maintain a balance between the gas being pumped out and the gas being pumped in. The laser power was controlled in order to maintain the temperature of the chondrule.

The experiments were conducted for atmospheric condition and sub-atmospheric conditions (50mbar). The initial temperature of the chondrule was maintained at 2400K in all the experiments. Three different cooling rates were set to compare the results, 100K/s, 200K/s and 400K/s. The time vs. temperature curves were recorded for all the experiments. Images of chondrule at the corresponding times were also recorded. All the diagnostics, viz. pyrometer and high speed camera were synchronised with the time of the clock installed in the HTCNL system to establish a correlation. A select set of crystalline chondrules were subject to SEM for surface morphology imaging. The observations and their interpretation will follow in the next sections.

## 3. Results and Discussion

Two sets of experiments were conducted – one, under atmospheric pressure and the other at a pressure of 50mbar. The processes of crystallisation and recalescence were observed and recorded in terms of temperature vs. time data and time-stamped images of the chondrule before, after and during the process of recalescence. Select chondrules from both the sets of experiments were scanned under SEM to understand the evolution of surface morphology under different pressure and cooling conditions. A detailed discussion on the effect of cooling rate, on the evolution of the surface features has been presented in this section. A vis-à-vis comparison based on the temperature-time history and evolution of surface morphologies in the two cases has also been presented. The entire section has been divided in four sub-sections, viz. A) Comparison based on temperature-time-history, B) Discussion on atmospheric experiments, C) Discussion on sub-atmospheric experiments, and D) Comparison based on surface morphologies..

**A) Comparison based on temperature-time-history**

Forsterite (chemical formula; name) is known to have its melting point at 2163 K [Reference]. In the present experiments the forsterite chondrules were heated to a temperature of 2400 K, which is ~250 K higher than the melting point. Then the chondrules were let to cool at different rates. Interestingly, all the samples were observed to have hypercooling before the onset of the crystallization process, irrespective of the cooling rate applied, as may be seen from Figure 3. The samples, irrespective of the pressure conditions, were observed to undergo recalescence from the melts; which is in proper agreement with the observed levels of undercooling from the work reported by Tangeman et al. [16] and Nagashima et al**. **[14]**.** Hypercooling refers to the undercooling of the melt after the limit of undercooling ($\Delta T_{hyp}$) is reached, which happens to be 425K for forsterite.[23].

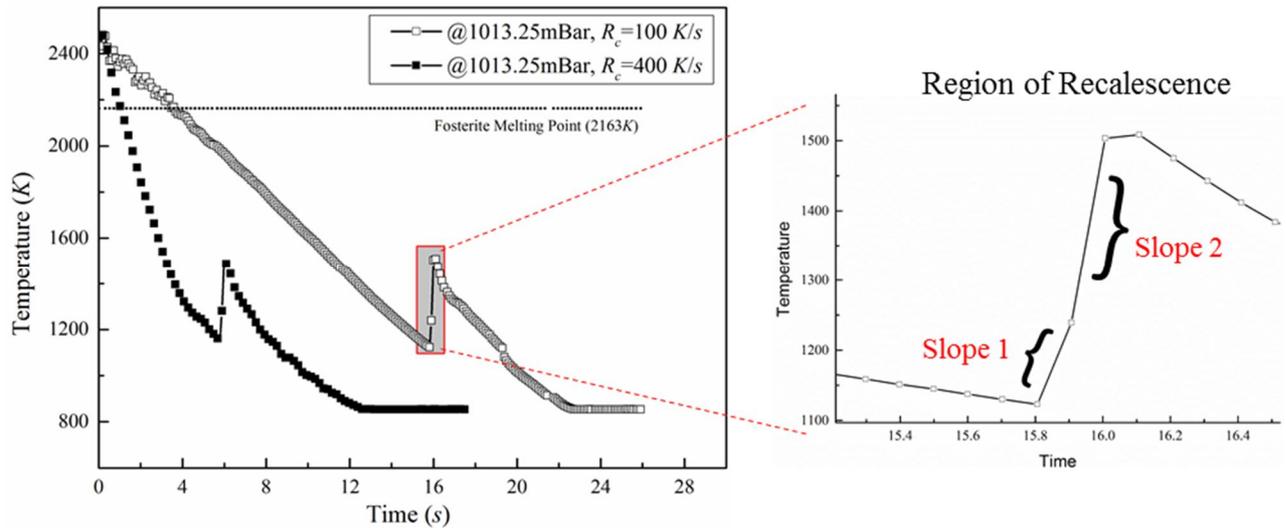

Figure 3 Temperature-time history of forsterite chondrules cooled at 100*K/s* and 400*K/s*.

The cooling curve of two identical forsterite samples which were let to cool down at 100*K/s* and 400*K/s*, have been shown in Figure 3. It can be seen that, for the cooling rate of 400*K/s*, which is four times as high as compared to 100*K/s*, the cooling time is less by the same factor, approximately. At a lower cooling rate, the temperature gradient between the melt's surface and body may be expected to be less. Thus, any sort of instabilities within the melt due to the temperature gradient is insignificant and hence, homogenous crystallisation may be expected. This is expected to lead to a relatively higher undercooling of the melt chondrule as compared to a case when cooling rate is high. However the observed difference is very less (~50K); which accounts for just 5% of undercooling that the samples undergo before crystallization. This is clearly observed from the averaged values of the undercooled temperature and degree of undercooling presented in Table 1 for different cooling rates (the averaging has been done after considering data from sufficient number of experiments carried out under the same conditions). As discussed earlier, the crystallization is seen to occur from hypercooled melts which is corroborated from the data given in Table 1.. However, the effect of difference in cooling rate on chondrule is observed during the recalescence process when recorded by the high speed camera.

Table 1: Averaged undercooled temperature and degree of undercooling for various cooling rates under atmospheric conditions.

| Cooling Rate (K/s) | Undercooled Temperature (K) | Degree of Undercooling (K) |
|---|---|---|
| 100 | 1180.7 | 982.29 |
| 200 | 1124.32 | 1038.68 |
| 400 | 1132.62 | 1030.37 |

The zoomed-in section of Figure 3 shows the temperature-time history of the chondrule while it undergoes crystallization. The sample rate of the pyrometer (10Hz) was fast enough to capture at least three data points within the recalescence region. These three data points give two different slopes which have been considered while comparing the cases of different cooling rates.

The data from the pyrometer has been used to estimate the rate of change of surface temperature with time. As the cooling process starts the fall in temperature (before the recalescence) depends only on the rate of cooling, however, during the process of recalescence (which is characteristic of homogenous nucleation), the heat of crystallisation is released which leads to a sudden temperature gradient between the surface of the chondrule and the surrounding atmosphere (Ar in this case). Therefore, the rate of change of temperature is expected to depend on the inner dynamics of the melt as well, along with the cooling rate. However, within the range of uncertainties involved, it is safe to state that the gradient is independent of the cooling rate employed.

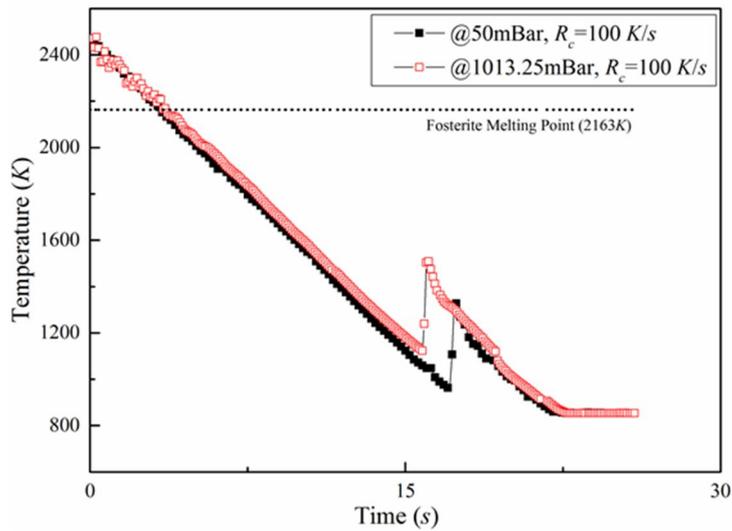

Figure 4: Cooling curve of the chondrules at cooling rate of 100*K/s* under both atmospheric and sub-atmospheric (50mBar) conditions.

Figure 4 represents the temperature-time history of two separate chondrules undergoing cooling at a rate of 100*K/s* but at two different pressure conditions, viz. atmospheric pressure and at a pressure of 50mBar. It is clearly seen that there is no difference in the pre-recalescence phase, as both the chondrules are subject to the same rate of cooling. However, the chondrule under atmospheric conditions was observed to approach the limit of undercooling earlier than the chondrule under sub-atmospheric condition. Thus, the onset of nucleation is delayed for the chondrule under sub-atmospheric conditions.

Also, under atmospheric conditions, there is a constant temperature gradient associated with the chondrule surface and the ambient gas, which causes convective plumes around the levitated chondrule. The channels of heat rejection during the process of undercooling are radiation, forced convective cooling by the levitating jet and the natural convection that is setup between the chondrule and the ambient gas. The conditions, however, are different when the sub-atmospheric conditions prevail, the radiative cooling remains nearly the owing to the fact that cooling starts from the same temperature in both the cases; as observed even in Figure 4. Since the flow rate of the levitating gas is lower in case of sub-atmospheric conditions, as discussed earlier, a reduction in the forced convective cooling of the chondrule occurs. Similarly, due to the rarefied ambient gas density the convective cooling gets subdued. This suppression of the convective cooling reduces the temperature gradient between the surface of the melt and its center which prolongs the sustenance of the melt and delays the onset of crystallisation process. This is imminent in Figure 5, where a comparison of the undercooling achieved and the degree of supercooling ($\Delta T_{undercooling}$) for the cases of atmospheric and sub-atmospheric conditions have been shown. The degree of supercooling is defined as the difference of the melting point temperature of the chondrule and the temperature of the onset of nucleation.

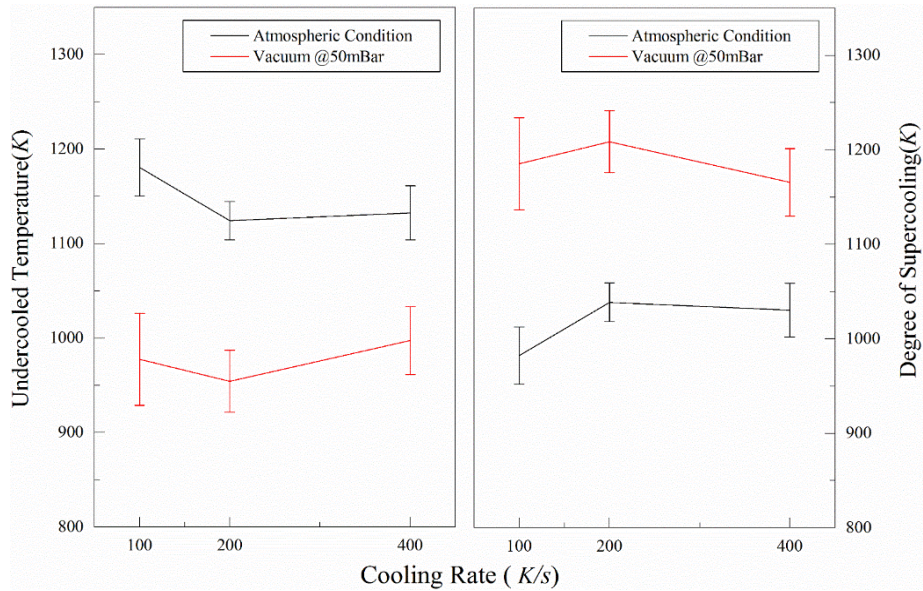

Figure 5: Comparison of undercooled temperature and the degree of supercooling for different cooling rates under atmospheric and sub-atmospheric conditions.

As it is seen from Figure 5, there is significant difference in the degree of supercooling between experiments under atmospheric conditions and sub-atmospheric conditions. This is attributed to the delay in the onset of nucleation process (and hence crystallisation) due to the factors accounted above. It may also be seen that the cooling curve is independent of the cooling rate employed, in both the cases. The degrees of supercooling are nearly equal, considering the errors that are involved, for both the cooling rates employed under atmospheric and sub-atmospheric conditions. This is owing to the fact that effect of temperature homogeneity obtained under slower cooling rates, say 100$K/s$, is not significant enough to cause a delay in the onset of the nucleation process. In contrast to this, the factors originating due to the difference in ambient pressures are strong enough to cause a delay in the nucleation process leading to observable differences in the cooling curves. Similar observations have been reported by Nagashima et al. [14] where the authors had compared the nucleation temperature of enstatite , while heterogeneously crystallizing it using a sample holder, under normal atmospheric conditions and under microgravity conditions. The authors reported a delay in the onset of nucleation for experiments under microgravity conditions and attributed it to the apparent suppression of convection. In a similar microgravity experiment, authors [22] have reported that $CaCO_3$ solution a was found to be more susceptible towards homogenous nucleation because of the lack of convection within the solution and hence difficulties were faced in achieving heterogenous nucleation in $CaCO_3$ solution.

Under the sub-atmospheric conditions, $\Delta T_{undercooling}$ is higher (~150K – 200K) as compared to that under the atmospheric condition and hence, a finer grain structure is expected for melts crystallized in the former case. Blander et al. [24] have observed that, with the increase in the magnitude of supercooling, the morphological trend of the crystals shifted from bar dendrites to fibres to sub-microscopic structures. Although they had conducted experiments with a mixture of forsterite and enstatite melts, and the setup was in a non-levitated environment which had led to heterogenous nucleation during crystallization. The range of supercooling achieved by the authors was 400K – 800K; which in our case, both under atmospheric and sub-atmospheric conditions, are higher. Hence, it is more likely that even under atmospheric conditions, fine sub-microscopic features may be obtained in our experimental conditions.

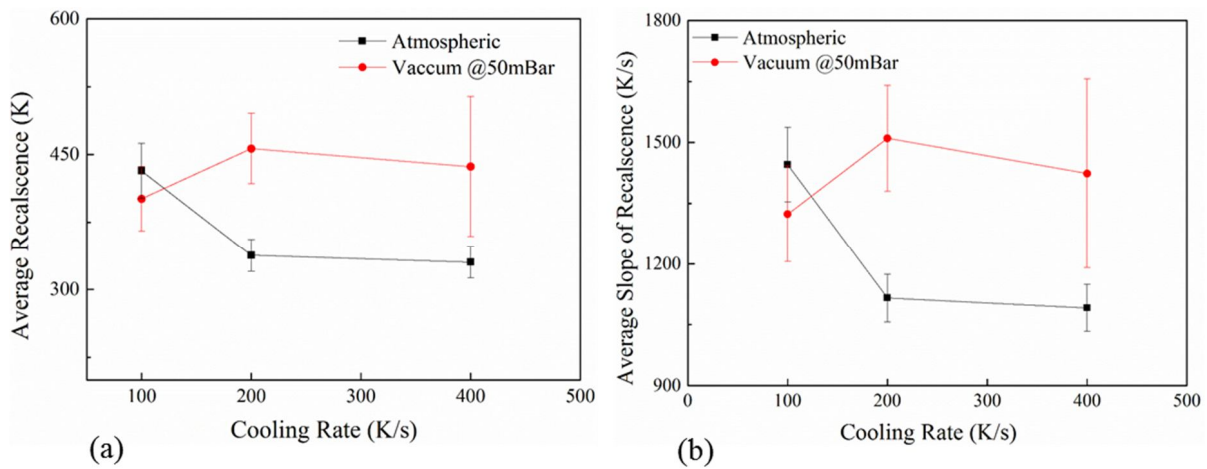

Figure 6: (a) The average amount of recalescence obtained in experiments; (b) The average slope of recalescence obtained for various cooling rates.

Figure 6 a and b show the degree of recalescence and the average slope of the recalescence process obtained in the experiments, respectively. The average slope of recalescence is obtained by taking the average of the two slopes observed in the temperature-time history of each chondrule under experimental conditions. It can be clearly seen from Figure 6 (a) that there is a remarkable difference in the magnitude of recalescence between sub-atmospheric and atmospheric conditions. Similar trend is also observed in Figure 6 (b). Under sub-atmospheric conditions, the chondrule tends to get undercooled to a larger degree and in the process, its crystallization is delayed, as discussed earlier. The difference in the amount of recalescence observed between atmospheric and sub-atmospheric conditions may be explained as follows. The molecules of the melt under sub-atmospheric conditions would have larger intermolecular distance, as compared to the atmospheric case, due to the lesser pressure in the former case. Thus during recalescence, the molecules of the melt have to traverse an more in order to solidify and in the process, the melts release the extra amount of energy which is reflected as the difference in the recalescence.. The same is also, observed in the average slope of the recalescence. Even if there is higher degree of recalescence in the sub-atmospheric case, the melt appears to crystallize within the same time period as in the atmospheric case (Figure 4 for reference); giving rise to higher slope in recalescence. This observation also implies the fact that melt undergoes a faster crystallization process under sub-atmospheric conditions and hence, one can expect finer surface features.

However, higher recalescence under sub-atmospheric condition for a cooling rate of 100$K/s$ is not evident from figures 6 a and b. The reason for this is in-built in the experimental procedure. It has been observed that, under sub-atmospheric conditions, the melt chondrule gets slightly more levitated, as compared to the atmospheric condition. . Similarly, while getting cooled down, the melt shifts downwards, coming closer to the nozzle. This shift is larger for slower cooling rates. This shift is seen to take place in the pre-recalescence phase. As, described earlier the pyrometer is focussed at the surface of the chondrule and it records the temperature based on the optical radiation. So, it can be understood that the sudden shift in the position of the chondrule makes it practically impossible for the pyrometer to record the temperature of the same place while heating and cooling. It is due to this shift that, for the cooling rate of 100$K/s$, the pyrometer is unable to capture the rise in the temperature during recalescence. The authors believe that, this experimental limitation prohibits the observation of any significant difference in the degree of recalescence for a cooling rate of 100$K/s$ The same reasoning explains the trend observed in case of the slope of recalescence.

The unusual trend where the degree of recalescence and, inadvertently, the slope of recalescence is higher by a very large margin for the applied cooling rate of 100$K/s$, than the rest under atmospheric conditions, may again be understood by the way in which the experiments were carried out. . The entire process of cooling or heating depends on the power delivered by the laser, which is the only sourse of heat. The laser power, on the other hand depends on a feedback mechanism, which has the pyrometer at its core. As the pyrometer sees the surface temperature, it gives a feedback to the laser, so as to achieve a desired cooling or heating rate. Thus, for a cooling rate of 100$K/s$, the laser power would decrease very slowly over time as compared to 400$K/s$ . . Thus, in case of cooling rate of 100$K/s$, the melt is exposed to laser for a longer duration during (even during recalescence). Laten heat released during recalescence summed up with this extra laser power accounts for a significant increase in the amount of recalescence (in terms of observed temperature) obtained at a cooling rate of 100K/s when compared to the rest of the cooling rates. However, for sub-atmospheric conditions, this is not observed due to the down shift of the melt during cooling, as discussed earlier.

**B) Discussion on atmospheric experiments**

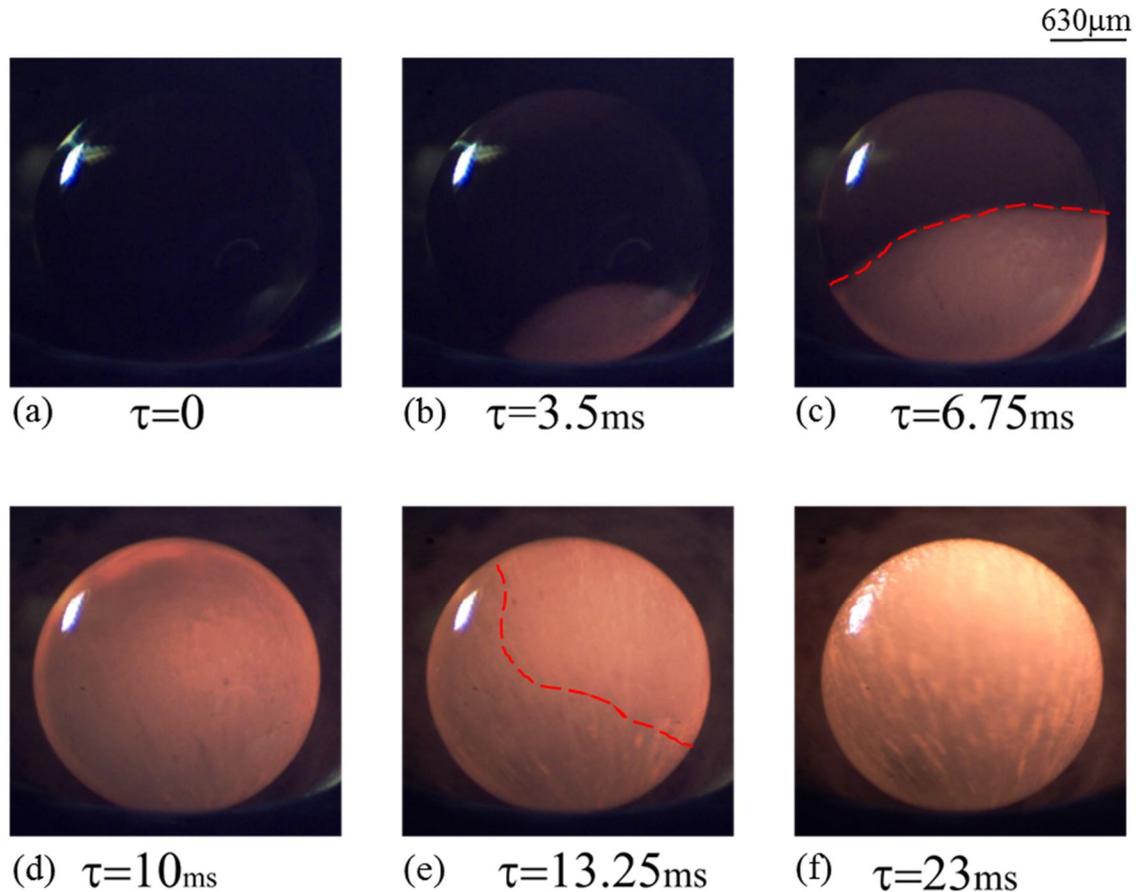

Figure 7: Time-sequenced images of the recalescence process for a sample cooled at 100*K/s* under atmospheric conditions.

Figures 7 and 8 show the time-stamped HSC images of the chondrule undergoing recalescence under atmospheric conditions with a cooling rate of 100*K/s* and 400*K/s* respectively. A small bulge, as seen in the figure is an entrapped gas bubble, which might have caught-up during the process of melting. This bubble might be thought of as a local inhomogeneity, serving as a nucleation center, but looking at the clear ring-shaped solid front advancing with time keeps that possibility out of consideration.

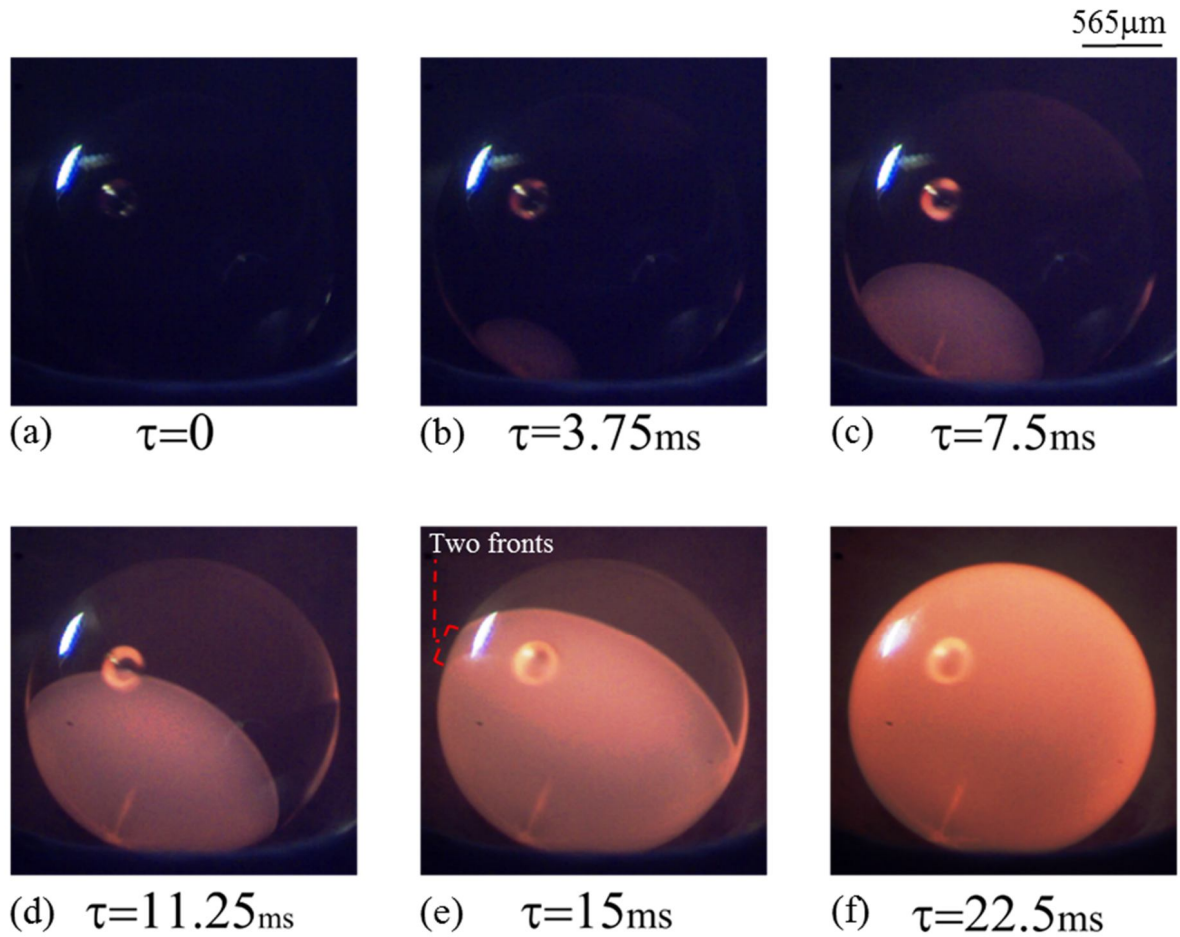

Figure 8: Time-sequenced images of the recalescence process for a sample cooled at 400K/s under atmospheric conditions.

Comparing the images depicted in Figure 7 and 8, it becomes evident that the cooling patters are different. In clear contrast to cooling at 400K/s resulting in clear rim-like solidification front, cooling at 100K/s has a diffused rim-like front. While the cooling rate is high (400K/s), there exists a high temperature gradient between the surface and the ambient and hence the surface cooling dominates, which also means that there is temperature inhomogeneity between the surface and the interior of the melt. Due to the surface cooling faster than the interior, a sharp rim-like solidification front is observed. Whereas at a lower rate of cooling (100K/s), the entire melt chondrule cools down as a whole and the temperature distribution across the entire volume is homogenous . The prevailing temperature conditions, in this case, allows the melt chondrule to crystallize from the core to the surface. This idea is well corroborated from the recalescence images in Figure7, where along the diffused rim, a radial growth of the crystal structures from the point of origin of the nucleation is seen. Similar growth pattern was observed by Miura et al. [20], when they carried out a simulation to study the crystallisation of a hypercooled melt under heterogenous nucleating conditions. A quasi-planar growth interface with solid growing radially from the seeding point was reported. However, in our case, the radial growth and the formation of a diffused rim were observed, simultaneously. A more careful observation reveals that, at a higher cooling rate the solidification front maintains the curvature of the rim-like structure from the onset of recalescence to the end. Whereas, in case of a lower cooling rate the semi-circular shape of the front tends to be flat as the time elapses. These observations are also in agreement with the results reported by Miura et al. [20]. Apart from this, one can further notice two sets of growth fronts during the recalescence process in both the cases. A strong black front is accompanied by a faded black front, as marked separately in Figure 8(e). However in Figure 7 (refer to 7(c) and 7(e)), the faded black front is rather much delayed. Since there are two types of cooling occurring in a melt, the surface cooling can be realted to the strong black front while the volumetric cooling can be referred as the faded black front. For higher cooling rate, the volumetric cooling alongwith the surface cooling is on higher side. This suggests that the strong temperature gradient developed in higher cooling rate cases not only aids in faster surface crystallization but, also accelerates the volumetric crystallization. However, in slow cooling rate cases, the gradient inside the melt is not that strong; due to which the volumetric crystallization is delayed. As observed in Figure 7, the faded black front is no more in definite shape, as like in Figure 8, but rather a patch that eventually fades out representing the end of recalescence process. This trait and behaviour of the chondrule during recalescence is also seen in sub-atmospheric conditions.

With the images shown in Figure 7 and, it is possible to make a qualitative assessment of the surface features that may have formed while cooling the chondrules at different rates. as it has been discussed that for high cooling rate, there is predominance of surface cooling under atmospheric conditions, which is rapid in nature. The surface of the solidified melt, therefore, appears to be smooth and clear However, for the case of lower cooling rate, volumetric cooling is predominant and there is a likelihood of crystallization occurring from within the melt and then reaching outward to the surface. From Figure 7 it is clearly seen that there are stripes that are oriented parallelly and pointing away from the nucleation point. These distinctions became clearer with the SEM images and the same have been shown in Figure 9. SEM electron micrographs have been shown only for those samples, which have been shown in images so as to ensure proper correlation between images and electron micrographs. This scheme has been adhered to throughout the article.

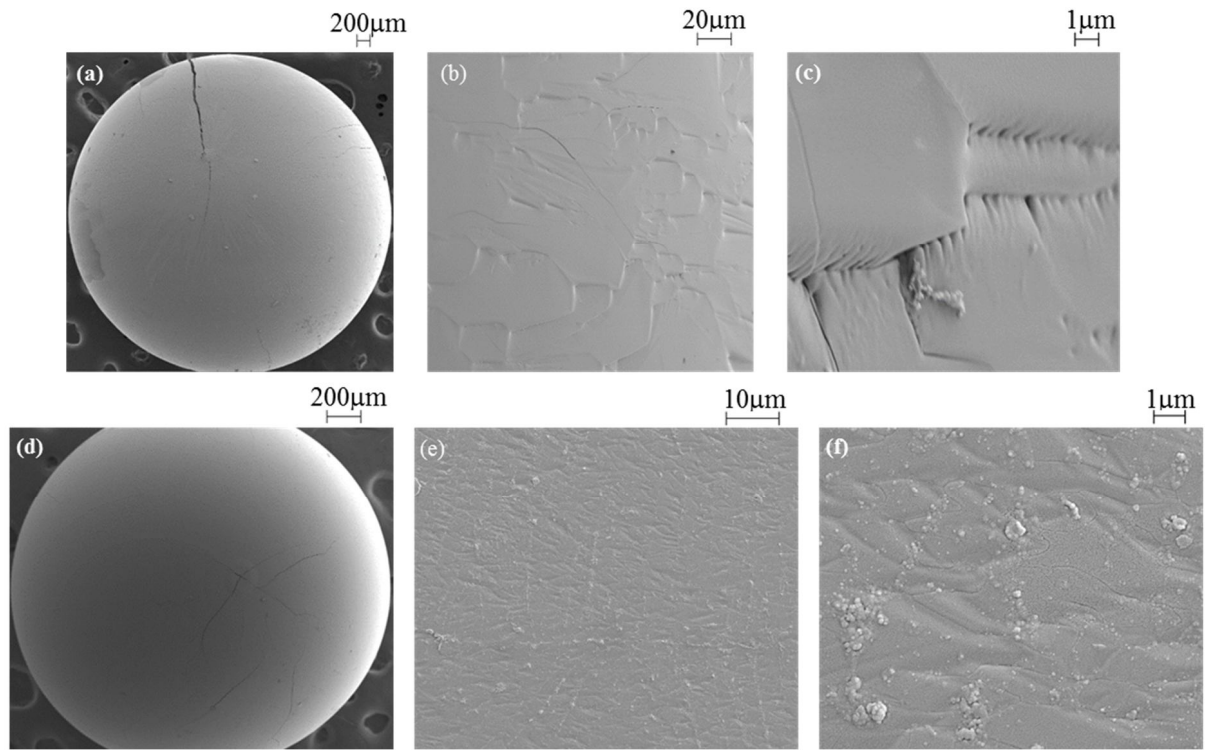

Figure 9: SEM images of the surface morphologies of the chondrules crystallized under atmospheric conditions with a cooling rate of 100*K/s* (a, b, c) and 400*K/s* (d, e, f).

Figures 9(a, b, c) show the electron micrographs of the surface features for low cooling(100K/s) , and Figures 9(d, e, f) show that for the high cooling rate (400K/s). As discussed earlier, based on the observations from the high speed camera images, it is clearly seen that larger surface textures are present on the surface of the chondrule that was cooled at 100K/s, while the surface features are pretty small in the other case. These observations may be explained on the basis of cooling time, that the chondrules get in both the conditions. When the cooling rate is lower the melt gets enough time to aggregate and form larger crystals and hence bigger surface features are observed. On the other hand, rapid cooling leads to the formation of finer structures and a smooth textured surface, because there is very little time available for aggregation before the melt solidifies. Similar results were reported by Kaushik et.al.[25], albeit in the case of cooling of plasma plume resulting in the formation of metallic nanoparticles. There it was found that prolonged cooling of plasma plume resulted in the formation of larger-sized nanoparticles, whereas rapid cooling under same conditions led to the formation of smaller-sized nanoparticles.

## C) Discussion on sub-atmospheric experiments

Under sub=atmospheric conditions, the melt chondrule undergoes larger supercooling and recalescence, as discussed above. This is primarily because of the suppression of convection and hence minimal thermal fluctuations. Also, as established from Figure 6(b), crystallization process is sped up in sub-atmospheric conditions and hence, one can expect finer structures on crystallisation and hence a smoother texture of the surface of the solidified chondrule, when compared to samples cooled under atmospheric conditions.

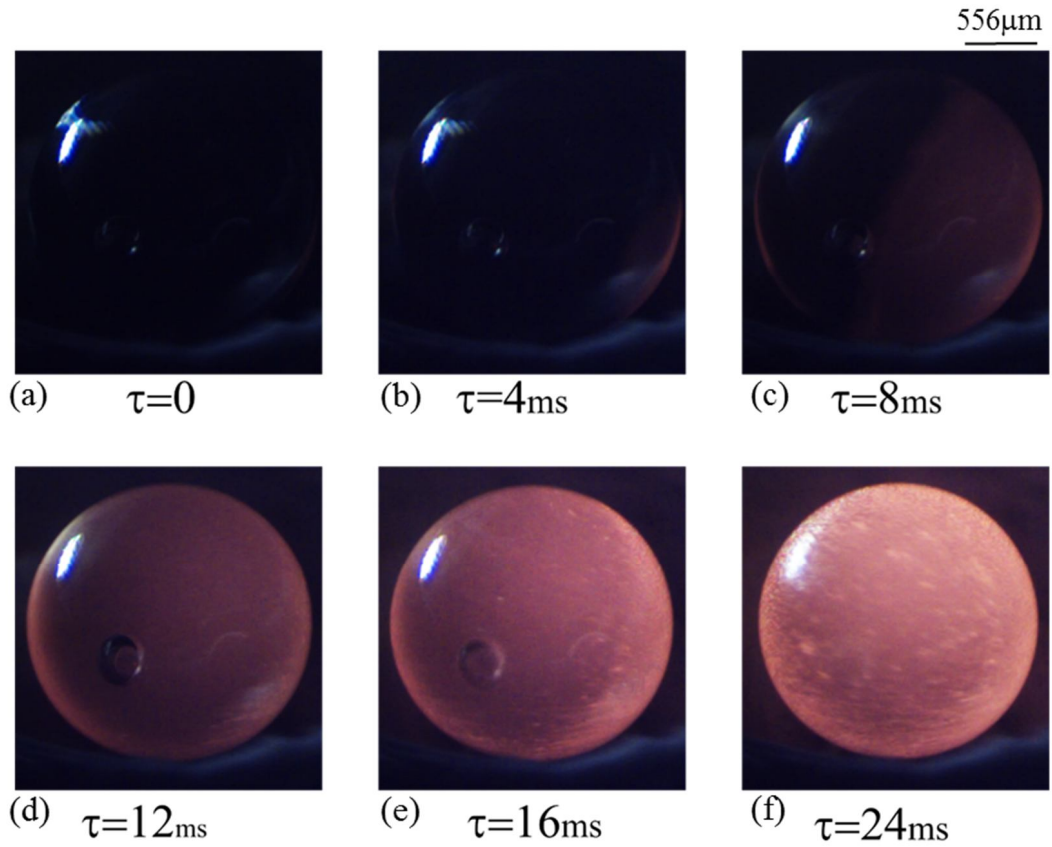

Figure 10: Time-stamped recalescence images of the chondrule subjected to a cooling rate of 100K/s under 50mBar sub-atmospheric conditions.

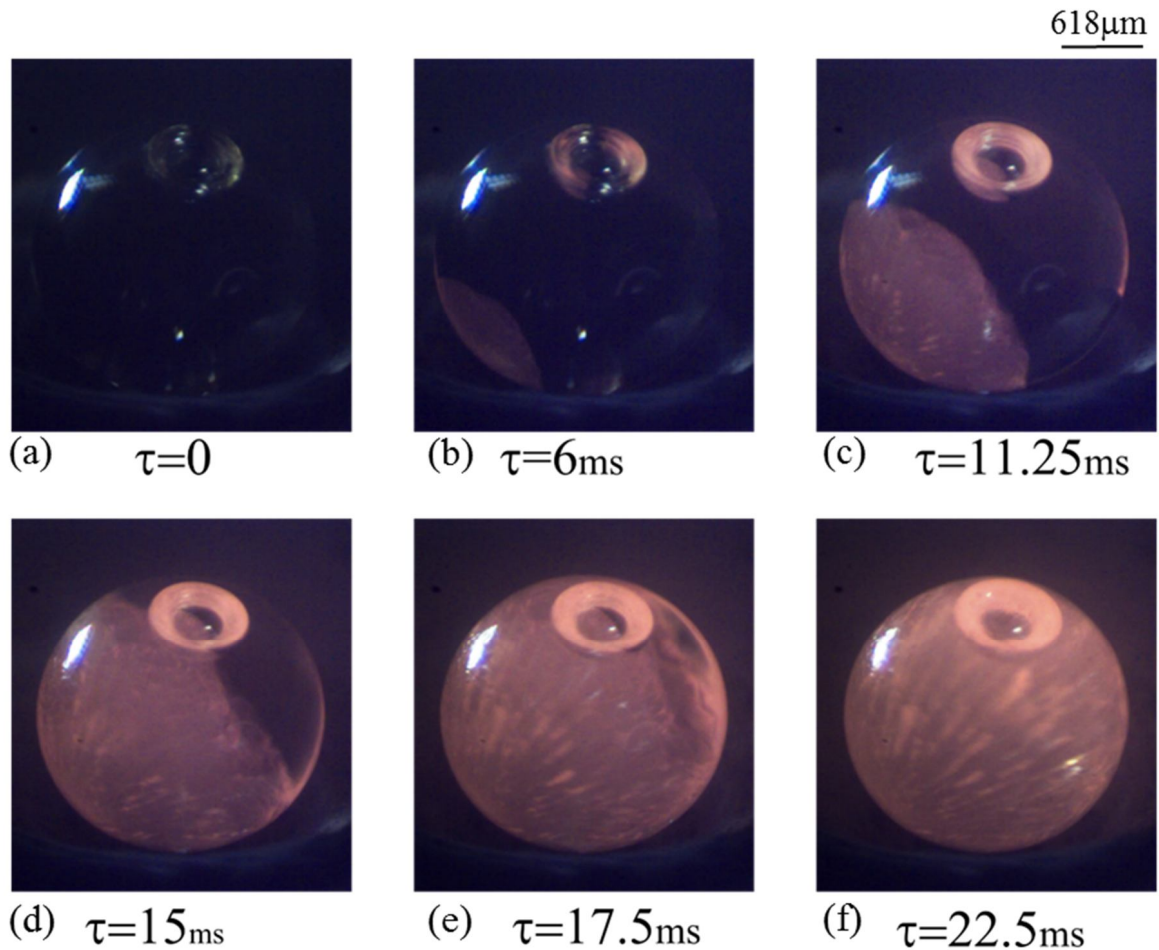

Figure 11: Time-stamped recalescence images of the chondrule subjected to a cooling rate of 400*K/s* under 50mBar sub-atmospheric conditions.

Figure 10 and 11 represents the time-sequenced images of the melts undergoing crystallization under 50mBar pressure conditions with cooling rates of 100K/s and 400K/s respectively. It is evident from the comparison of the two figures that there is a diffused front visible in the case of lower cooling rate.

As discussed earlier, at the cooling rate of 110K/s and lower pressure of 50mBar, volumetric crystallisation is dominant and this is clearly observed in Figure 10(f). Here clear striped patches are visible and are distributed all over the solidified melt surface. These stripped patches are randomly distributed but are arranged parallel and pointing away from the nucleation center; implying the fact that the crystal structure may be quasi-planar as was also observed and reported by Miura et al. [20].

As expected in case of higher rate of cooling, the volumetric crystal front is seen to be rapidly following the surface crystallization front and the same is observed from Figure 11 b and c. Another salient observation, is that the nucleation is not influenced by the presence of surface inhomogeneities such as air bubble. Under sub-atmospheric conditions, though major cooling occurs through the surface, yet cooling of the interior of the melt closely follows the surface cooling owing to the fact that the temperature inhomogeneity inside the melt is minimum. The cooccurrence of the surface and volume crystallisation may be observed in figures 11 c, d and e. In Figure 11(c), slight patches are visible which get developed into long-striped patches as seen in Figure 11(e) as the crystal front spreads over the volume of the chondrule. Similar to the patches discussed in the case of slow cooling rate these patches are aligned away from the nucleation center. Although the solidification fronts in Figure 11, seem to be strikingly similar to rimmed growth fronts that have been shown in Figure 8, yet a closer look suggests that the edges of the crystal front are broken and irregularly shaped, unlike what has been seen in Figure 8. The growth pattern of the crystal front observed from Figure 11 is quite similar to the numerically projected growth pattern front in [20]. From Figure 11(e), the surface front is observed to have concluded earlier than the volumetric front, suggesting that although both the fronts initiate and grow simultaneously, yet the volumetric crystallization takes a little longer to conclude. A discussion on the same supported with the experimental observations will follow in the next subsection.

Figures 12 a, b and c are the electron micrographs of the surface features of the chondrule cooled at 100K/s. while figures d, e and f are for the chondrule cooled at 400K/s.

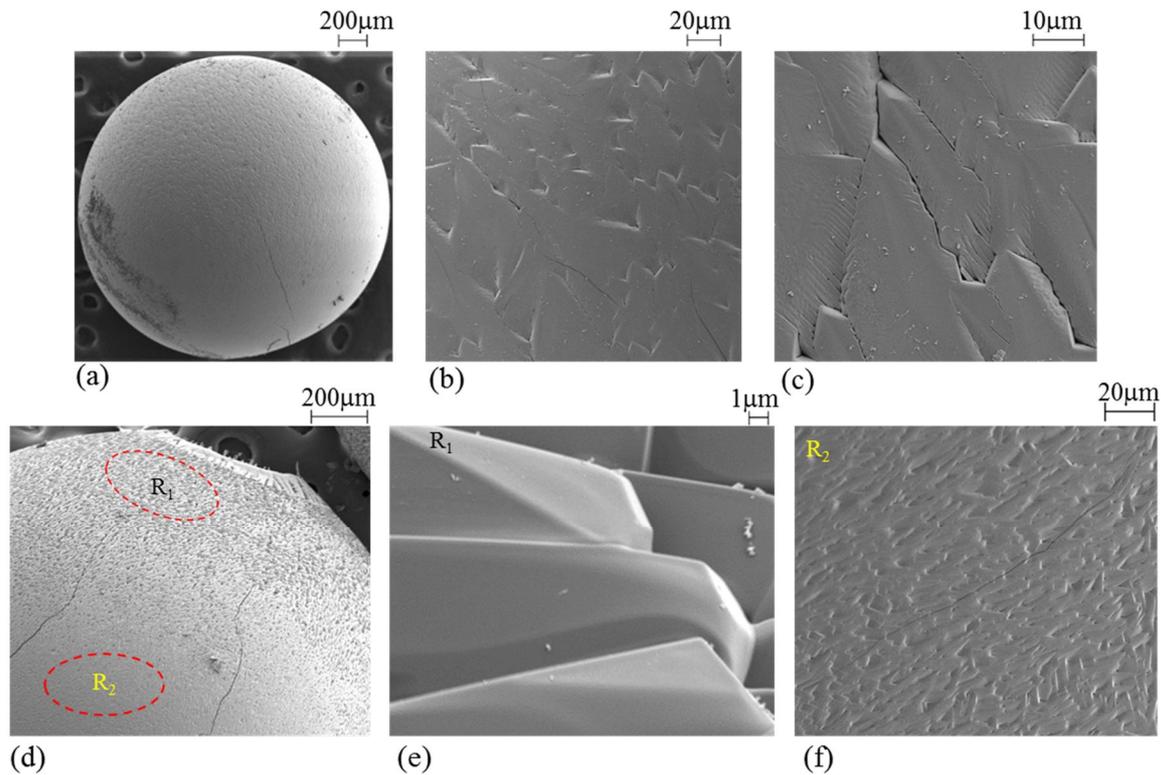

Figure 12: SEM images of the surface morphologies for the samples crystallized under 50mBar sub-atmospheric conditions with a cooling rate of 100*K/s* (a, b, c) and 400*K/s* (d, e, f).

In case of lower cooling rate there are similar kind of features that are seen to be distributed over the surface. Contrary to the case of lower cooling rate, in case of higher cooling rate, two clearly distinct regions (R1 and R2 as indicated in Figure 12 d) can be seen. The region R1 which is located in the close proximity of the gas bubble is seen to be composed of block-like structures. On the other hand, in region R2 which is located far away from the gas bubble is seen to have irregular depressions. . The difference in the microstructures as observed in Figure 12(d) can be explained by considering the dynamics of crystallized grains in region R1 immediately after the removal of gases entrapped within the bubble. As it may be seen from Figure 11f, the entrapped gas bubble (seen in Figure 11(f)) gets ejected from the crystallized solid surface following recalescence. This ejection causes dislocation of the granules near the bubble region and pushes them outward. The dislocation of the granules appear as small blocks stacked one above another, as observed in Figure 12(e). Figure 13 shows the zoomed-in electron micrograph of the transition that takes place between regions R1 and R2 as shown in Figure 12(d). It clearly shows that the region R1 has dominant structures made up of block-like stacked structures, whereas the region R2 comprises of relatively smoother surface textures. This goes without saying that surface of chondrule cooled at a higher rate has a smoother texture with finer features, compared to its counterpart that was cooled at a lower rate.

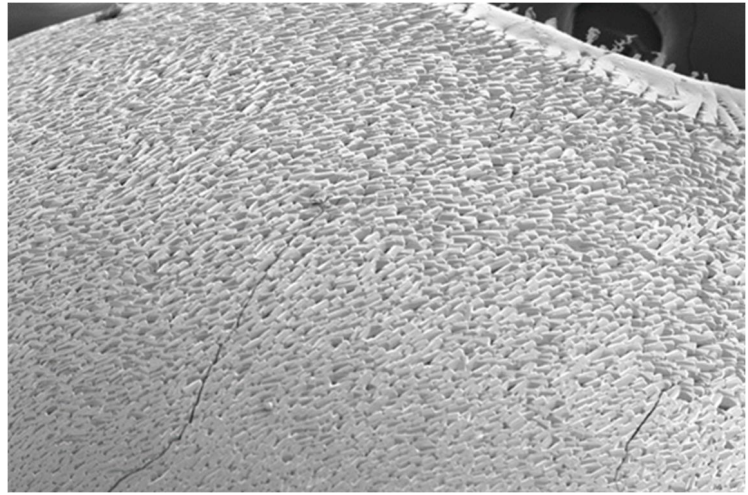

Figure 13: A zoomed-in view of the transition in the surface morphology of the sample crystallized under 50mBar sub-atmospheric conditions with a cooling rate of 400*K/s*

## D) Comparison based on surface morphologies

Salient differences in the temperature data obtained under atmospheric and sub-atmospheric conditions have been discussed in the preceding sub-sections. It has been found that in case of the melt chondrule crystallising under sub-atmospheric condition the undercooling and recalescence both were higher. Also, a comparative study of the surface features and the texture have been presented this far, for both atmospheric and sub-atmospheric conditions and for different cooling rates. However, no direct comparison of the scale of the surface features have been made and this is what is going to be discussed in this sub-section with the help of Figure 16.

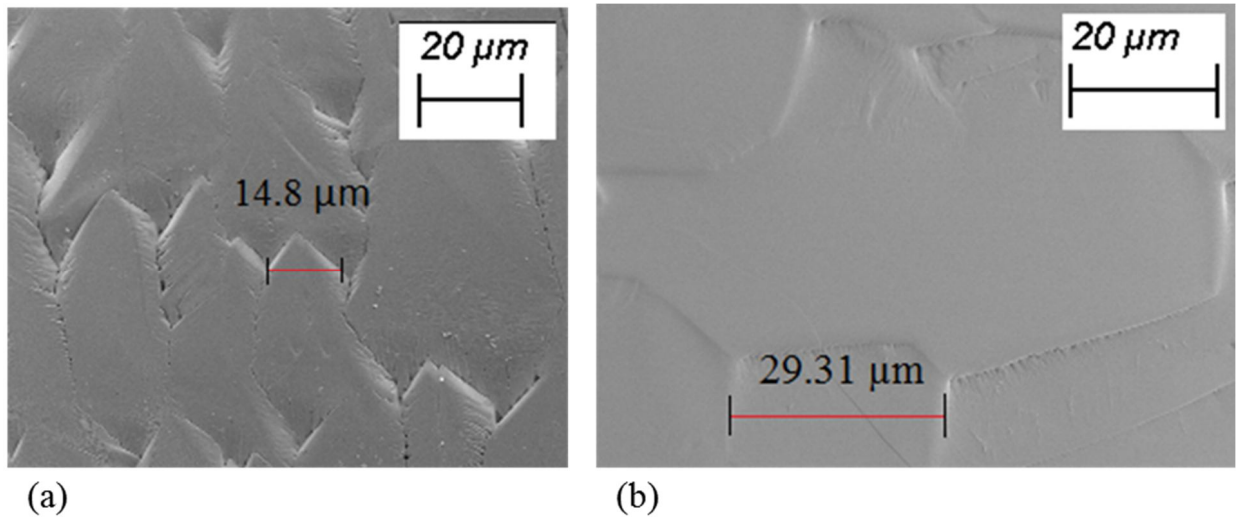

Figure 16: Comparison of the surface textures of forsterite melts crystallized with a cooling rate of 100K/s under sub-atmospheric conditions (a) and atmospheric conditions (b).

Figures 16 a and b show the electron micrographs of the surface of the chondrules cooled under sub-atmospheric condition at a rate of 100K/s. It may be clearly seen that due to faster rate of cooling post-recalescence, the features developed on the surface of the chondrule that was cooled under sub-atmospheric conditions are smaller compared to the features observed on its counterpart cooled under atmospheric condition. This corroborates the results reported by researchers that faster cooling and higher extent of undercooling lead to finer surface structures [24, 25]. The authors would like to highlight the fact that the images shown in Figure 16 are representative and are in proper agreement with the statistical trend observed from many such images taken at various locations of the sample surfaces. Only two images and the cooling rate of 100K/s are being discussed for the sake of brevity and avoid internal repetition of results as similar results have been obtained in the case of 400K/s and from images of various locations of the same chondrules.

## Special Case

Having discussed all the possible cases and the outcomes of the experiments, in this section the authors are presenting a special case that is a direct experimental proof of the numerical simulations reported by Miura et. al. and to give an insight of the evolution of two solidification fronts, viz. surface and volumetric. The case relates to the cooling of the chondrule at 400K/s at an ambient pressure of 50mBar. Figure 14 shows the time-tagged images recorded during the cooling process.

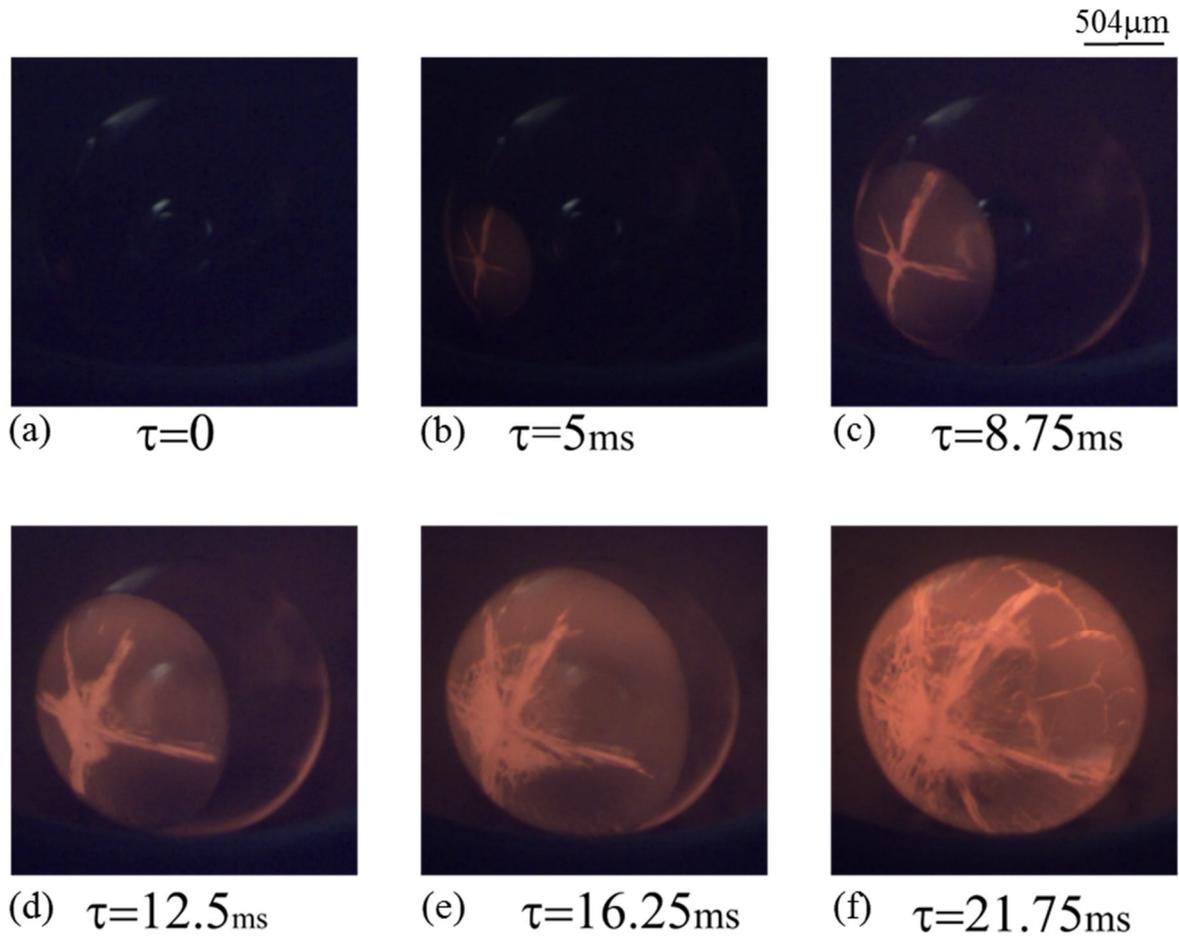

Figure 14: Time-stamped recalescence images of the chondrule subjected to a cooling rate of 400*K/s* under 50mBar sub-atmospheric condiions

In Figure 14, starting from (a) to (f), the dendritic arms and the rim both are visible. At an earlier instant, as shown in in Figure 14(b) both, the rim and the dendrite can be seen to have grown to the same extent, where the tip (farthest extent) of the dendritic arm (formed due to volumetric cooling) just touches the perimeter of the rim (formed due to the solidification of the surface), suggesting that both are having the same velocities. However, the path of travel is different for either one of them. Rim grows on the chondrule surface, while the dendrite grows through the center of the chondrule to the outer surface.

Over the period of time, both grow at competitive pace and in Figure 14(d) one of the arms of the dendrite is still seen to be able to match the speed of the growth of the rim. But as the crystallization continues, the dendritic growth is seen to have slowed down and the distance between the rim and the dendrite is observed to be growing, as is seen in Figure 14(e).

The reason for the same may be explained as follows. The rim grows along the surface and its rate is controlled by the temperature gradient between the surface and the ambient, which in-turn depends on the externally applied cooling rate. Since the cooling rate applied is a constant, the growth of the rim is also constant and uninhibited. However, the growth of dendrite results from cooling over the volume, which means that it will always have a dependence on the temperature gradient between the surface and interior of the chondrule. Under any given condition there will be a gradient existing between the center and the surface of the chondrule, center being at higher temperature. This temperature gradient will impede the growth of the dendrite which starts form the surface and starts moving towards the center. Although it has been experimentally found and discussed in the preceding subsections that the gradient is less in the sub-atmospheric conditions, still it is found to be strong enough to affect the dendritic growth. Although slowed down, yet the dendritic growth continues and it reaches the center where the temperature is highest. The difference between this temperature and the temperature at the surface is what checks the speed of the dendritic growth. As the dendrite growth crosses the center, its growth speed picks up as the temperature starts decreasing from the center to the surface and hence, the resistance also begins to decrease from the center to the surface. The rim growth gets completed by the time the dendritic growth reaches the other side of the chondrule surface. The completion of the recalescence process can be seen in Figure 14(f) where all the dendritic arms are observed to have spread all over the volume of chondrule. The electron micrograph of the surface morphology of the chondrule has been shown in Figure 15. Close observations show dendritic structures (black circle) on the surface and also re-establish the fact that due to cooling under sub-atmospheric conditions, the textures on the surface are fine.

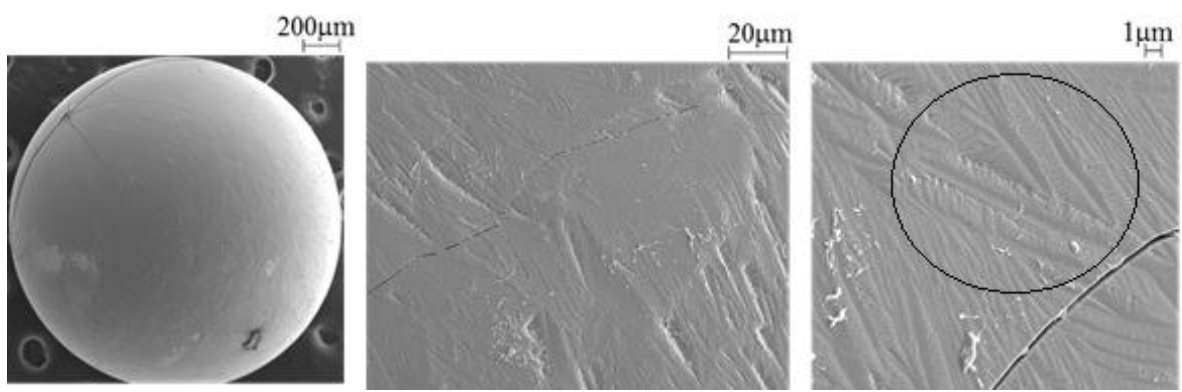

Figure 15: SEM images of the surface morphologies for the samples crystallized under 50mBar sub-atmospheric conditions with a cooling rate of 400K/s

The images shown in in Figure 14 were discussed here because these results were numerically predicted by Miura et al. [20] for a supercooled forsterite melt under heterogeneously seeded conditions. The major difference of those numerically simulated results with the present study is that the this study is pertaining to the homogenous crystallization wherein the forsterite melt has been hypercooled and the applied cooling rate is two to three times lesser than what was applied in the numerical study. Even though the whole mechanism of initiating the crystallization in our study, and the factors governing it, are different from the reported numerical study, yet the growth mechanism observed has been found to be in good agreement with the simulated results. This also suggests that the growth mechanism need not be unique to a case and may well be well suited to explain different cases.

## 4. Conclusion

Experiments were performed to study the crystallisation process of the forsterite samples melt and maintained at 2400K and let to cool thereafter at three different cooling rates (100K/s, 200K/s and 400K/s) and at different pressures (atmospheric pressure and 50mBar), in Argon atmosphere. It was found that melts crystallized under sub-atmospheric conditions attained higher degree of undercooling, when compared to their counterparts at atmospheric pressure. Further, it was observed that even if the initial conditions remain same the recalescence is stronger in case of sub-atmospheric condition and so is the cooling rate. This also led to the finding that the surface morphologies are different for the chondrules cooled under different conditions – the ones that were cooled under atmospheric condition had larger surface features as compared to those cooled in sub-atmospheric conditions. The extra heat generated during the recalescence in the sub-atmospheric conditions were explained using the theory of extra traversal of molecules at a lower pressure. The authors have also presented the images from the high-speed camera and tried to explain the process of rim-like solid front in case of surface cooling and dendritic structures in case of volume cooling and the reason behind the difference in their growth rates. As per the best of the authors' knowledge the experimental observation of contact-less solidification of the molten forsterite chondrules under the sub-atmospheric condition have been reported for the first time and the results thus obtained and presented in this article are found to be in good agreement with the numerical results that are present in the literature.